\documentclass[11pt]{amsart}
\usepackage{amsfonts,amssymb,amsmath}
\usepackage{latexsym}
\usepackage{color}
\usepackage{graphicx}
\usepackage{epsfig}
\usepackage{hyperref}
\usepackage{bm}

\usepackage[active]{srcltx}

\newtheorem{theorem}{Theorem}[section]

\newtheorem{example}[theorem]{\sf Example}

\newtheorem{case}[theorem]{\sf Case}

\newcommand {\C}    {{\mathbb C}}
\newcommand {\D}    {{\mathbb D}}

\newcommand {\R}    {{\mathbb R}}
\newcommand {\N}    {{\mathbb N}}
\newcommand {\Z}    {{\mathbb Z}}

\newcommand {\ag}    {\mathfrak{a}}
\newcommand {\bg}    {\mathfrak{b}}

\newcommand {\0}    {|\,0\rangle}

\newcommand{\al}{\alpha}
\newcommand{\pa}{\partial}

\newcommand{\la}{\lambda}
\newcommand{\ta}{\tau}

\newcommand{\de}{\delta}

\newcommand{\rar}{\rightarrow}

\newcommand{\non}{\nonumber}

\begin{document}

\date{\today}

\title[Spectral theory in the Fock space]{On the spectral theory in the Fock space\\
with polynomial eigenfunctions
}
\author{Alexander V Turbiner}
\address{UNAM, Mexico-city, Mexico\\ (corresponding author)}
\email{turbiner@nucleares.unam.mx}
\author {Nikolai L Vasilevski (deceased)}
\address{Cinvestav, Mexico-city, Mexico}

\begin{abstract}
The notion of the eigenvalue problem in the Fock space with polynomial eigenfunctions 
is introduced. This problem is classified by using the finite-dimensional representations 
of the $\mathfrak{sl}(2)$-algebra in Fock space.
In the complex representation of the 3-dimensional Heisenberg algebra,
proposed in \cite{TV}, this construction is reduced to the linear differential operators
in $(\frac{\pa}{\pa \overline{z}}\,,\,\frac{\pa}{\pa z})$ acting on the space 
of poly-analytic functions in $(z,\overline{z})$. The number operator, equivalently, 
the Euler-Cartan operator appears as fundamental, it is studied in detail. The notion of (quasi)-exactly solvable operators is introduced. The particular examples of the Hermite and Laguerre operators 
in Fock space are proposed as well as the Heun, Lame and sextic QES polynomial operators.    

\end{abstract}

\maketitle

\section{Introduction}

In \cite{Turbiner:1995} the spectral problem 
\[
    L(b,a)\, \phi(b) \, \0\ =\ \la\, \phi(b)\, \0 \ ,
\]
was introduced, see also \cite{Turbiner:2016-PR}, where 
\[
    L(b,a) \ =\ \sum A_{i,j} b^i a^j \ ,
\]
is a finite-degree polynomial defined as the sum of ordered monomials in two non-commutative variables $(b,a)$ with vacuum $\0$, subject to the condition $a \0 = 0$, $\phi(b)$ is an eigensolution with eigenvalue $\la$. It was shown that if $L(b,a)$ is a particular polynomial of second degree in $a$, denoted as
\[
   L_e(b,a)\ =\ P_2(b) a^2 + P_1(b) a + P_0     \ ,
\]
where $P_2, P_1, P_0$ are polynomials in $b$ of the second, first and zeroth degrees, 
respectively, and $b,a$ are elements of the three-dimensional Heisenberg algebra $\mathbb{H}_{3}(b,a,I)$ with $[b,a]=-I$, the spectral problem has infinitely-many 
polynomial eigenfunctions,
\[
 \phi_n\ =\ p_n(b)\ ,\ 
\] 
of degrees $n=0,1,2,\ldots\ $.
Such an operator $L_e(b,a)$ was called {\it exactly-solvable}.
In particular, in the coordinate-momentum representation, $b=x, a=\pa_x$ and $\0=1$, 
where $\pa_x \equiv \frac{d}{dx}$ is the derivative, 
the operator $L_e$ becomes the familiar hypergeometric operator and $p_n(b)$ are nothing but 
the hypergeometric polynomials. 
It is evident that $L_e(b,a)$ can be rewritten in terms of $(ba)$ and $a$ which span the maximal affine subalgebra $b_2$ of the $\mathfrak{sl}(2)$-algebra: $[ba, a] = -a$. Furthermore, by taking a more general operator
\[
   L_q(b,a)\ =\ Q_4(b) a^2 + Q_3(b) a + Q_2     \ ,
\]
where $Q_{4,3,2}(b)$ are polynomials of the fourth, third and second degrees, respectively, and imposing two conditions on the coefficients in front of the leading and sub-leading terms of the coefficient functions $Q$, respectively, $(n+1)$ polynomial eigenfunctions of the form
\[
 \phi_n\ =\ q_n(b)\ ,
\]
occur, while all other eigenfunctions are not finite degree polynomials. In this case the operator $L_q$ can be rewritten in terms of the generators of the algebra $\mathfrak{sl}(2)$ 
in its $(n+1)$-dimensional representation realized in terms of $(b,a)$, see below. Such an operator was called 
{\it quasi-exactly-solvable}. In the coordinate-momentum representation, $b=x, a=\pa_x, \0=1$, 
the operator $L_q$ becomes the Heun operator $L_q(x,\pa_x)=P_4(x) \pa_x^2 + P_3(x) \pa_x + P_2(x)$. 
In general, there are no other than $L_{e,q}$ operators of second degree in $a$ admitting 
polynomial eigenfunctions.

The Heisenberg algebra $\mathbb{H}_{3}$ can be realized in terms of differential, difference \cite{t-sl22} and discrete operators \cite{CT}. By substituting these representations into $L_{e,q}$ we arrive at the differential, difference and discrete operators with the property of polynomial isospectrality: they admit polynomial eigenfunctions with eigenvalues which do not depend on the type of operator. 
The goal of the present work is to study the operators $L_{e,q}$ when they act on the complex plane \cite{TV} as differential operators in $(z,\overline{z})$.
Their eigenspaces consist of poly-analytic functions with eigenvalues of infinite multiplicity. We are going to study three operators: the number operator (Section 2), the exactly-solvable operator (Section 3), the quasi-exactly-solvable operator (Section 4). The exposition follows with minimal changes the last (tenth) draft prepared by the second author NLV before his passing away. 
Personal recollections by AVT about the second author - recently deceased 
Professor Nikolai L Vasilevski - conclude the article. 

The paper is somewhat incomplete: in the absence of the second author (NLV) the first author (AVT) 
did not take the responsibility of presenting proofs in order to make it rigorous. This paper is the result of the collaboration of a professional theoretical physicist and a professional mathematician.

\section{Extended Fock construction and the number operator}

Let us introduce the three-dimensional Heisenberg algebra $\mathbb{H}_{3} = \{ \ag,\bg,1 \}$ with commutator $[\ag, \bg]\ = \ 1$  and $[\ag, 1]=[\bg, 1]=0$. It is worth noting that the generator $\ag$ acts as derivative:
\[
 [\ag,\bg^n]=n \bg^{n-1}\ .
\]

A formal construction of the Fock space, see e.g. \cite[Chapter 5, Section 5.2]{BerezinShubin} is based on two formally adjoint operators $\ag$ and $\bg = \ag^{\dag}$ defined on a dense linear subset of a separable Hilbert space $\mathcal{H}$, where they act invariantly and satisfy the commutation relation  $[\ag, \ag^{\dag}]\ = \ I$. Furthermore, there is also a normalized element $\0 :=\Phi_0 \in \mathcal{H}$, $\|\Phi_0\|=1$, called the vacuum vector, such that $\ag \Phi_0 =0$, and the linear span of elements $(\ag^{\dag})^n\Phi_0$ with $n \in \mathbb{Z}_+$ is dense in~$\mathcal{H}$, for discussion see \cite{TV}.

We will deal with a more general situation where, first, the operator $\bg$ is not necessarily formally adjoint to $\ag$ and, second, $\dim\ker\ag \geq 1$.

Our extended Fock space construction, see e.g. \cite{Vas_Prepr-2022}, is given by the representation of the algebra $\mathbb{H}_3$ in a separable Hilbert space $\mathcal{H}$. This implies that there are two operators $\ag$ and $\bg$ 
defined on a dense linear subset of a separable Hilbert space $\mathcal{H}$, where they act invariantly and satisfy the commutation relation $[\ag, \bg]\ = \ I$. Furthermore,
the linear subset $\ker \ag := L_{[1]} $ spanned by the vacuum vectors, is non-trivial 
with $\dim L_{[1]} \geq 1$;
and the set, formed by the finite linear combinations of elements from the linear subsets $L_{[n]} := \bg^{n-1} L_{[1]}$, $n \in \mathbb{N}$, is dense in $\mathcal{H}$.

The Hilbert space $\mathcal{H}$ possessing the above properties is naturally called 
the \emph{Fock space related to  $\ag$ and $\bg$}, or just the \emph{$(\ag,\bg)$-Fock space}.

Note that the classical Fock formalism corresponds to the case when $\bg = \ag^{\dag}$ is formally adjoint to $\ag$, and the space $L_{[1]}$  is one-dimensional, generated by a single element $|0\rangle :=\Phi_0 \in \mathcal{H}$.

Let us make two remarks to this definition. First, the operators $\ag$ and $\bg$ generate in fact the representation  of the universal enveloping algebra $U_{\mathbb{H}_{3}}$ of $\mathbb{H}_3$ in $\mathcal{H}$. And second, as examples show, the subspases $L_{[n]}$ may or may not be closed, as well as, they may or may not be mutually orthogonal to each other.

One of the most important elements of the universal enveloping algebra $U_{\mathbb{H}_{3}}$ is
\begin{equation}
\label{L0}
   {\sf L}_0\ =\ \bg \ag    \ .
\end{equation}
Usually, it is called the number operator and represents the Euler-Cartan generator $J^0_1$ of the $\mathfrak{sl}(2)$-algebra \cite{t-sl22}, {which for all $k \in \C$ can be generated by}
\begin{eqnarray}
\label{sl2ag}
   J^+_k &=&  \bg^2 \ag - k\, \bg\ , \non \\
   J^0_k &=&  \bg \ag - \textstyle{\frac{k}{2}}\ , \\
   J^-_k &=& \ag \non\ .
\end{eqnarray}

Let us describe the action of the operator ${\sf L}_0$ on the elements of $L_{[n]}$: if we take $h_{[n]} = \bg^{n-1}h_{[1]} \in L_{[n]}$, where $h_{[1]} \in L_{[n]}$, then
\begin{equation*}
 {\sf L}_0 h_{[n]} = \bg(\ag\bg^{n-1}h_{[1]}) = \bg(n-1)\bg^{n-2}h_{[1]} = (n-1)h_{[n]}\ .
\end{equation*}
This means that the space $L_{[n]}$ is the eigenspace of the operator ${\sf L}_0$ with the corresponding eigenvalue $(n-1)$. Moreover, the operator $L_0$ is bounded on $L_{[n]}$, thus
the action
\begin{equation} 
\label{eq:action_L_0}
L_0|_{L_{[n]}} = (n-1)I: L_{[n]} \ \longrightarrow \ L_{[n]}
\end{equation}
can be extended by continuity to the closure $\overline{L_{[n]}}$, in case the space $L_{[n]}$ is not closed.

Note that the eigenvalues $\la_n = n-1$, $n \in \N$, are equidistant, and the lowest eigenspace is $\overline{L_{[1]}}$, the vacuum space, or stated differently, the space of vacuum vectors. Furthermore, in the case $\ker \bg = \{0\}$, the above eigenvalues exhaust the spectrum of the operator ${\sf L}_0$. At the same time, \cite[Subsection 3.2]{Vas_Prepr-2022} gives an example of operators $\ag$ and $\bg$ with non-trivial $\ker \bg$ in the corresponding Hilbert space, such that the spectrum of the operator ${\sf L}_0\ =\ \bg \ag$ coincides with $\Z$.

\medskip
Let us give several examples involving different realizations of the $(\ag,\bg)$-Fock spaces.


\begin{case} {\sf Coordinate-momentum representation} \\ 

{\rm
\begin{equation}
\label{c-m-reps}
   \ag\ =\ \frac{d}{dx}\ \mbox{and}\ \bg= x\ ,
\end{equation}
}
\end{case}

\begin{example}\label{ex:c-m} {\sf Coordinate-momentum representation over the real line} \\ 
{\rm
Let $\ag = \frac{d}{dx}$ and $\bg= x$, be defined on the linear space of all polynomials in $x$, which is dense in $\mathcal{H} = L_2(\R, e^{-x^2}dx)$.
The space $L_{[1]} = \ker \ag$ is one-dimensional, generated by the vacuum vector $|0\rangle  = \bm{1}$, the function identically equals to $1$ on $\R$. Then each space $L_{[n]}$ is also one-dimensional, generated by $x^{n-1}$, $n \in \N$, the eigenvalues of the operator ${\sf L}_0 = x\frac{d}{dx}$ are equal to $(n-1)$ with corresponding eigenvectors $x^{n-1}$, $n \in \N$.
}
\end{example}

\begin{example}\label{ex:delta}{\sf Coordinate-momentum representation over the uniform lattice} {\rm (see  \cite{Turbiner:1995} for details).\\
Introduce the so-called Norlund derivative (the shift, translation invariant derivative over the uniform lattice) $\ag_{\delta}$, \ $\delta \in \R \setminus \{0\}$, related to the so-called {\it umbral calculus} by Gian-Carlo Rota, and the operator $\bg_{\delta}$ as follows
\begin{equation}
\label{delta}
        \ag_{\delta} f(x) = \frac{f(x + \de) - f(x)}{\de}\ , \qquad \bg_{\delta} f(x) = x f(x-\delta)\ ,
\end{equation}
densely defined on the linear space of all polynomials in $x$ over $\mathcal{H} = L_2(\R, e^{-x^2}dx)$. The space 
$L_{[1]} = \ker \ag_{\delta}$ is one-dimensional, generated by the vacuum vector $|0\rangle  = \bm{1}$. Then each space $L_{[n]}$ is also one-dimensional and is generated by $x^{(n-1)}$,
where $x^{(n-1)}=x(x-\de)(x-2\de)\ldots (x-(n-2)\de)$ is the so-called {\it quasi-monomial}. The eigenvalues of the operator ${\sf L}_0 =x\,\frac{1-e^{-\de \pa_x}}{\de} $ are $(n-1)$ with corresponding eigenspaces $L_{[n]}$, $n \in \N$.
}
\end{example}

\begin{example}\label{ex:q}{\sf Coordinate-momentum representation on the exponential lattice} 
{\rm (see e.g. \cite{CT} for details).\\
On the dense linear space of all polynomials in $x$ over $\mathcal{H} = L_2(\R, e^{-x^2}dx)$ we introduce the dilation operator $T_qf(x) = f(qx)$, \ $q \in \R_+$, and then
\begin{equation}
\label{q}
 \ag_q = x^{-1} \frac{T_q -1}{q-1}\quad , \quad \bg_q = x\frac{d }{dx}(q-1)(T_q - 1)^{-1}x\ .
\end{equation}
Note that the operator $(T_q - 1)^{-1}$ is well-defined on elements of the form ``$x \times \, polynomial$'', and that the operators $\ag_q$ and $\bg_q$ act on monomials $x^n$ as follows
\begin{equation*}
 \ag_q x^n = \{n\}_qx^{n-1}, \qquad
 \bg_q x^n = \frac{n+1}{\{n+1\}_q} x^{n+1},
\end{equation*}
where $\{n\}_q =\frac{q^n -1}{q-1} = q^{n-1} + q^{n-2}+ \ldots +1$ is the so-called $q$-number.

Note that $\ag_q$ is the so-called Jackson derivative (the discrete, dilation-invariant derivative over the exponential lattice), related to the so-called {\it quantum calculus} by Stanislaw L Woronowicz \cite{W:1989}. The space $L_{[1]} = \ker \ag_{q}$ is one-dimensional, generated by the vacuum vector $|0\rangle  = \bm{1}$. Then each space $L_{[n]}$ is also one-dimensional, generated by
\begin{equation*}
 \bg_q^{n-1}\bm{1} = \frac{(n-1)!}{\{2\}_q \cdots \{n-1\}_q} x^{n-1}.
\end{equation*}
Observe that the number operator ${\sf L}_0 = x\frac{d}{dx}$ does not depend on $q$, its eigenvalues are $(n-1)$ with corresponding eigenspaces $L_{[n]}$, $n \in \N$.
 }
\end{example}

Note that the operators $\ag_{\delta}$ and $\bg_{\delta}$ of Example \ref{ex:delta}, as well as the operators $\ag_q$ and $\bg_q$ of Example \ref{ex:q} become the operators $\ag$ and $\bg$ of Example \ref{ex:c-m} in the limit $\delta \to +0$ and $q \to +1$, respectively.

\smallskip
In the next three examples the operators $\ag$ and $\bg$ are densely defined over the span of the subspaces $L_{[n]} = \bg^{n-1}\ker \ag$ of the corresponding Hilbert space $\mathcal{H}$.

\begin{example}{\sf The harmonic oscillator} 
{\rm (see e.g. \cite[Sections 1.8 and 5.2]{BerezinShubin} for details).\\

\begin{equation}
\label{c-a-reps}
  \ag = \frac{1}{\sqrt{2}}\left(x+ \frac{d }{d x}\right)\ \mbox{and} \ \bg = \ag^\dag = \frac{1}{\sqrt{2}}\left(x - 
  \frac{d }{d x}\right)
\end{equation} 
be the standard annihilation and creation operators, densely defined in  $\mathcal{H} = L_2(\R)$.
The space $L_{[1]} = \ker \ag$ is one-dimensional, generated by the vacuum vector $|0\rangle  = e^{-\frac{x^2}{2}}$. Then each space $L_{[n]}$ is also one-dimensional, generated by $H_{n-1}(x)e^{-\frac{x^2}{2}}$, where $H_{n}(x)$ are Hermite polynomials, $n \in \N$. The spectrum of the Harmonic oscillator
\begin{equation}
\label{Harm-Osc}
    H = \frac{1}{2} \left(\frac{d^2 }{d x^2} + x^2\right) = \ag^{\dag}\ag +\frac{1}{2} = {\sf L}_0 +\frac{1}{2}
\end{equation}
consists of the eigenvalues $\lambda_n = n-\frac{1}{2}$, whose corresponding eigenvectors are 
$H_{n-1}(x)e^{-\frac{x^2}{2}}$, $n \in \N$. Hence, the number operator coincides with the Hamiltonian of the harmonic oscillator. 
}
\end{example}

\begin{example}{\sf The unit disk $\D$ case} {\rm (see  \cite[Subsection 3.1]{Vas_Prepr-2022} for details).\\
For each $\lambda > -1$, introduce the probability measure
\begin{equation}
\label{disk}
 d\mu_{\lambda} = (\lambda +1)\left(1 - z\overline{z}\right)^{\lambda}dA(z)\ , \quad \mathrm{where} \quad dA(z) = \frac{1}{\pi}dxdy\ ,
\end{equation}
and the Hilbert space $L_2(\mathbb{D}, d\mu_{\lambda})$, as well as the operators $\ag = \frac{\partial }{\partial \bar{z}}$ and $\bg = \bar{z}$ densely acting therein.
The space $L_{[1]} = \ker \ag$ is infinite-dimensional and coincides with the classical Bergman subspace $\mathcal{A}^2_{\lambda}(\D)$ of $L_2(\mathbb{D}, d\mu_{\lambda})$ which consists of the analytic in $\D$ functions. Then each space $L_{[n]}= \bar{z}^{n-1}\mathcal{A}^2_{\lambda}(\D)$ is also infinite-dimensional. The eigenvalues of the operator ${\sf L}_0 = \bar{z}\frac{\partial }{\partial \bar{z}} $ are $(n-1)$ with corresponding infinite-dimensional eigenspaces $L_{[n]}$, 
$n \in \N$ \footnote{This case will not be studied further in this paper.}.
 }
\end{example}

\begin{example}{\sf Landau magnetic Hamiltonian }{\rm (see \cite{Mouayn2011} and \cite{TV} for details).\\
Consider the operators 
\begin{equation}
\label{repr-in-z}
  \ag = \frac{\pa}{\pa \overline{z}}\ \mbox{and}\ \bg = \ag^\dag = -\frac{\partial}{\partial z} + \overline{z}
\end{equation}
densely defined in $\mathcal{H} = L_2(\mathbb{C},d \mu)$, with the Gaussian measure $d\mu(z)= \frac{1}{\pi}e^{-z\cdot {\bar z}}dxdy$. The space $L_{[1]} = \ker \ag$ is infinite-dimensional and coincides with the classical Fock subspace $F^2(\C)$ of $L_2(\mathbb{C},d \mu)$ which consists of the analytic in $\C$ functions. Then each space $L_{[n]}= (-\frac{\partial}{\partial z} + \overline{z})^{n-1}F^2(\C)$ is also infinite-dimensional and coincides with the true-$n$-poly-Fock space $F^2_{(n)}(\C)$, see \cite{Vasilev00,TV} for details. In this case the number operator ${\sf  L}_0$ coincides with the Landau magnetic Hamiltonian:
\begin{equation}
\label{landau}
  {\sf L}_0 = - \frac{\pa^2}{\pa z\,\pa\overline{z}} + \overline{z}\,\frac{\pa}{\pa\overline{z}}\ ,
\end{equation}
which (in suitable units and up to an additive constant) is a realization in the Hilbert space $L_{2}(\C,d\mu)$ of the similarity-transformed Schr\"odinger operator (which we call here the {\it Landau magnetic Hamiltonian}) describing the transverse motion of a charged particle evolving in the complex plane $\mathbb{C}$ subject to a normal uniform constant magnetic field in the asymmetric (Landau) gauge, see e.g. \cite{LL:1977}, Chapter XV, $\S$112. The first term in ${\sf L}_0$ has a meaning of the kinetic energy. The spectrum of ${\sf L}_0$ consists of infinitely many equidistant eigenvalues, each of infinite multiplicity (Landau levels), they are of the form $\lambda_n = n-1$ and the corresponding eigenspaces are nothing but the true-$n$-poly-Fock spaces $F_{(n)}^2(\mathbb{C})$. 
 }
\end{example}


We consider now spectral problems for the elements of the subalgebra $Alg({\sf L}_0)$ of the
universal enveloping algebra $U_{\mathbb{H}_{3}}$ generated by ${\sf L}_0$. Each element of $Alg({\sf L}_0)$ is nothing but the polynomial
\begin{equation*}
 \mathcal{P}({\sf L}_0) = \sum_{j=0}^k c_j{\sf L}_0^j,
\end{equation*}
where $k$ depends on the polynomial $\mathcal{P}$ in question. Then \eqref{eq:action_L_0} implies that
\begin{equation*}
 \mathcal{P}({\sf L}_0)|_{L_{[n]}} = \mathcal{P}(n-1)I: L_{[n]} \ \longrightarrow \ L_{[n]},
\end{equation*}
which can be extended by continuity to the closure $\overline{L_{[n]}}$, in case the space 
$L_{[n]}$ is not closed.
This implies that the eigenvalues $\lambda_n$ of the operator $\mathcal{P}({\sf L}_0)$ are of the form $\lambda_n = \mathcal{P}(n-1)$, $n \in \N$,  and the lowest eigenspace is $\overline{L_{[1]}}$, 
which coincides with the vacuum space. Furthermore, in the case $\ker \bg = \{0\}$, it is evident the above eigenvalues exhaust the spectrum of the operator $\mathcal{P}({\sf L}_0)$.

\section{Exactly-solvable Problems in the Fock Space}

We consider here spectral problems for the elements of the bigger subalgebra 
$Alg({\sf L}_0,\ag)$ of the
universal enveloping algebra $U_{\mathbb{H}_{3}}$ generated by ${\sf L}_0$ and $\ag$. 
These elements span the maximal affine subalgebra of the $\mathfrak{sl}(2)$-algebra 
(\ref{sl2ag}) \cite{t-sl22}: $[\ag, {\sf L}_0]=\ag$. 
It is evident that the algebra $Alg({\sf L}_0,\ag)$ can be ordered: 
it is the algebra of the ordered monomials 
in $({\sf L}_0, \ag)$. Hence, each element of this algebra is a non commutative polynomial 
of ${\sf L}_0$ and $\ag$, which, by using the relation  $\ag {\sf L}_0 = {\sf L}_0 \ag + \ag$, 
can be uniquely represented 
in the form
\begin{equation}
\label{P}
 \mathcal{P}({\sf L}_0,\ag) = \sum_{i,j} c_{i,j} {\sf L}^i_0 \ag^j \ ,
\end{equation}
where $c_{i,j}$ are complex parameters. Note that the operator $\ag$ does not preserve any 
$L_{[n]}$: take $h_{[n]} = \bg^{n-1}h_{[1]} \in L_{[n]}$, where $h_{[1]} \in L_{[n]}$, then
\begin{equation*}
 \ag h_{[n]} = \ag\bg^{n-1}h_{[1]} = (n-1)\bg^{n-2}h_{[1]} \in L_{[n-1]} \ ,
\end{equation*}
thus, acting as a filtration. To fix this problem, for each $n \in\N$, we consider the subspaces 
$L_{[1]},\,L_{[2]}, \, \ldots \, , L_{[n]}$. Since they are linearly independent 
(\cite[Corollary 2.9]{Vas_Prepr-2022}), consider their direct sum
\begin{equation}
\label{curly L}
 \mathcal{L}_n = L_{[1]} \dotplus L_{[2]} \dotplus \ldots \dotplus L_{[n]}\ .
\end{equation}
Note that even if all subspaces $L_{[j]}$ are closed, their direct sum is not necessarily 
closed (see e.g. \cite[Example 3.2]{Vas_Prepr-2022}).

Furthermore, the family of nested (sub)spaces $\mathcal{L}_k$ forms an infinite flag in $\mathcal{H}$
\begin{equation}
\label{f}
 \mathcal{F}: \ \ \mathcal{L}_1 \ \subset \ \mathcal{L}_2  \ \subset \ \ldots \ \subset \ \mathcal{L}_n  \ \subset \ \ldots,
\end{equation}
and, as it is easily seen, each element $\mathcal{P}({\sf L}_0,\ag)$ preserves the flag $\mathcal{F}$, i.e.,  $\mathcal{P}({\sf L}_0,\ag)$ maps each subspace $\mathcal{L}_k$ into itself, and the space $\mathcal{H}$ can be recovered by the flag elements as
\begin{equation*}
 \mathcal{H} = \mathrm{closure} \left(\bigcup_{k \in \N} \mathcal{L}_k \right).
\end{equation*}

Observe that ${\sf L}_0 = J^0_0$ and $\ag = J^-_0$ of \eqref{sl2ag}, therefore the algebras $Alg({\sf L}_0,\ag)$ and $Alg(J^0_1, J^-_1)$ coincide.
We call the operators $\mathcal{P}({\sf L}_0,\ag)$ in the universal enveloping algebra $U_{\mathbb{H}_{3}}$, which preserve the flag $\mathcal{F}$, \emph{exactly-solvable}.

One of the important particular cases of  $\mathcal{P}({\sf L}_0,\ag)$ (\ref{P}) is represented by the second degree polynomial 
in $\ag$,
\begin{equation}
\label{P2}
 \mathcal{P}_2({\sf L}_0,\ag)\ =\  c_{2,0} {\sf L}^2_0 + {\sf L}_0 (c_{1,1} \ag + c_{1,0}) + c_{0,2} \ag^2 + c_{0,1} \ag + c_{0,0} \ ,
\end{equation}
or, by substituting ${\sf L}_0\ =\ \bg \ag $, equivalently, 
\begin{equation}
\label{P2ba}
\mathcal{Q}_2 (\bg,\ag)\ =\ P_2(\bg)\, \ag^2\ +\ P_1(\bg)\, \ag\ +\ P_0\ ,
\end{equation}
where $P_{2,1,0}(\bg)$ are polynomials of degrees 2,1,0 in $\bg$, respectively. In the coordinate-momentum representation 
$\ag = \frac{d}{dx}$ and $\bg= x$, the operator (\ref{P2ba}) becomes the standard hypergeometric (Riemann) operator,
\begin{equation}
\label{P2x}
\mathcal{P}_2 (x,\pa_x)\ =\ P_2(x)\, \pa_x^2\ +\ P_1(x)\, \pa_x\ +\ P_0\ ,
\end{equation}
It is evident that the eigenfunctions in the eigenspaces $\mathcal{L}_n$ (3.13) 
of (\ref{P2})-(\ref{P2ba}), (\ref{P2x}) are hypergeometric polynomials in $\bg$ and $x$, respectively. Explicitly,
\begin{equation}
\label{phi-n}
   \phi_n (\bg)\ =\ \sum^n h_i(\bg)^i\,\0 \in\ \mathcal{L}_n\  \ ,
\end{equation}
see (\ref{curly L}), where $h_i$ are parameters, and $h_n=1$ without loss of generality. The
corresponding eigenvalues are quadratic in the quantum number $n$, 
$$\la_n = c_{2,0}\,n(n-1) \,+\, c_{1,0}\,n\,+\,c_{0,0}\ . $$

In the complex $(z,\overline{z})$ representation (\ref{repr-in-z}) the operators ((\ref{P2}), (\ref{P2ba})) become fourth order differential operators,
\begin{equation}
\label{P2z}
 \mathcal{P}_2(z, \overline{z})\ =\  c_{2,0} ({-\pa_z \pa_{\overline{z}} + \overline{z}\pa_{\overline{z}}})^2 + 
 ({-\pa_z\pa_{\overline{z}} + \overline{z}\pa_{\overline{z}}}) (c_{1,1} \pa_{\overline{z}} + c_{1,0}) + c_{0,2} \pa_{\overline{z}}^2 + c_{0,1} \pa_{\overline{z}} + c_{0,0} \ ,
 \ 
\end{equation}
where $\pa_z \equiv \frac{\pa}{\pa z}\ ,\ \pa_{\overline{z}}\equiv\frac{\pa}{\pa \overline{z}}$. 
The discrete spectrum of this operator is infinite and is quadratic in the quantum number $n$. Needless to prove that the $n$th eigenspace is of infinite multiplicity and is 
given by an $n$-polyanalytic function. If $c_{2,0}=0$, the operator 
$\mathcal{P}_2(z, \overline{z})$ becomes a third order differential operator. If additionally $c_{1,1}=0$, it becomes a second order differential operator. In the special case $c_{2,0}=c_{1,1}=c_{0,2}=c_{0,1}=c_{0,0}=0$ and $c_{1,0} = 1$, then the operator 
$\mathcal{P}_2(z, \overline{z})$ coincides with the Landau magnetic Hamiltonian ${\sf L}_0$, 
see (\ref{landau}). 

Two particular cases of ((\ref{P2}), (\ref{P2ba})) should be mentioned.

\subsection{Hermite operator}

Let us take the expression
\begin{equation}
\label{P2hermite}
     \mathcal{P}_2 (\bg,\ag)\ =\ -\ag^2\ +\ \bg\, \ag\ .
\end{equation}
Its eigenvalues are equidistant, $\la_n=n-1,\ n=1,2,\ldots$. In the coordinate-momentum representation (\ref{c-m-reps}) $\ag = \pa_x\ \mbox{and}\ \bg= x\ $, 
the operator (\ref{P2hermite}) becomes the celebrated Hermite operator
$(-\pa_x^2 + x \pa_x)$ with eigenfunctions
\begin{equation}
\label{phi-n-hermite }
   \phi_n (\bg)\,\0\ =\ \sum^n h_i(\bg)^i\,\0\quad \rar \quad \phi_n(x) = \sum^n h^{(n)}_i x^i\ ,
\end{equation}
which are the Hermite polynomials, where $h_i^{(n)}$ are well-known coefficients. In the complex $(z,\overline{z})$ representation (\ref{repr-in-z}), the Hermite operator takes the form of a second order differential operator,
\begin{equation}
\label{hermite-zz}
     \mathcal{P}_2 (z, \overline{z})\ =\ -\pa_{\overline{z}}^2\ -\ \pa_z \pa_{\overline{z}} + 
     \overline{z}\pa_{\overline{z}}\ ,
\end{equation}
with the eigenfunctions/eigenspaces 
\begin{equation}
\label{phi-n-hermite-zz}
   \phi_n \ =\ \sum^n h_i^{(n)}\ (-\pa_z + \overline{z})^i\,\0 \ .
\end{equation}

\subsection{Laguerre operator}

Take the expression
\begin{equation}
\label{P2laguerre}
     \mathcal{P}_2 (\bg,\ag)\ =\ -\bg \ag^2\ +\ (\bg - \al-1)\, \ag\ ,
\end{equation}
where $\al$ is the index.
Its eigenvalues are equidistant, $\la_n=n-1,\ n=1,2,\ldots$. In the coordinate-momentum representation (\ref{c-m-reps})
$\ag = \pa_r\ \mbox{and}\ \bg= r\ $, the operator (\ref{P2hermite}) becomes the familiar Laguerre operator
$(-r\pa_r^2 + (r - \al-1) \pa_r)$ with the eigenfunctions
\begin{equation}
\label{phi-n-laguerre}
   \phi_n (\bg)\,\0\ =\ \sum^n l_i^{(n)}(\bg)^i\,\0\quad \rar \quad \phi_n(r)=\sum^n l^{(n)}_i r^i\ ,
\end{equation}
which are the Laguerre polynomials ${\tilde L}_n^{(\al)}$ with index $\al$, where $l_i^{(n)}$ are well-known coefficients. In the complex $(z,\overline{z})$ representation (\ref{repr-in-z}) the Laguerre operator takes the form of a third order differential operator,
\begin{equation}
\label{hermite-zz}
     \mathcal{P}_2 (z, \overline{z})\ =\ (\pa_z - \overline{z})\pa_{\overline{z}}^2\ -\ 
     ({\pa_z - \overline{z} + \al + 1)
     \pa_{\overline{z}}}\ .
\end{equation}
with eigenfunctions 
\begin{equation}
\label{phi-n-hermite-zz}
   \phi_n \ =\ \sum^n l_i^{(n)}\ (-\pa_z + \overline{z})^i\,\0 \ ,
\end{equation}

\section{Quasi-Exactly-Solvable Problems in the Fock Space}

Now let us take the universal enveloping algebra $U_{\mathfrak{sl}(2)}$ generated by the representation of the $\mathfrak{sl}(2)$-algebra,
\begin{eqnarray}
\label{sl2ag-L0}
   J^+_n &=&  \bg\, ({\sf L}_0 - n\,)\ , \non \\
   J^0_n &=&  {\sf L}_0 - \textstyle{\frac{n}{2}}\ , \\
   J^-_n &=& \ag \non\ ,
\end{eqnarray}
with $n \in \C$, cf.(\ref{sl2ag}), see \cite{Turbiner:1988,Turbiner:1994,t-sl22}, 
${\sf L}_0\ =\ \bg \ag $ (\ref{L0}).  
In this representation it is a subalgebra of the universal enveloping Heisenberg algebra, $U_{\mathfrak{sl}(2)} \subset U_{\mathbb{H}_{3}}$. 
It is evident that if $n$ is an integer, the $\mathfrak{sl}(2)$-algebra appears in {\it the type} 
of the finite-dimensional representation: the operators $J^+_n, J^0_n, J^-_n$ act in the space $\mathcal{L}_n$, see (\ref{curly L}), which coincides with $n$-poly-Fock space $F^2_{n}(\C)$, see \cite{TV} and also \cite{Vasilev00}, \cite{Vas_Prepr-2022} 
\footnote{We continue to call it the finite-dimensional representation}. 
A formal dimension of $F^2_{n}(\C)$ is $(n+1) \times \infty$: it is a polynomial in $\bar z$ 
of degree $n$ with analytic functions as coefficients.

We call the operators $\mathcal{P}({\sf L}_0,\ag,\bg)$ in the universal enveloping algebra $U_{\mathfrak{sl}(2)}$, which preserve the $n$-poly-Fock space $F^2_{n}(\C)$, \emph{quasi-exactly-solvable}. 
Among them, the most important operator is given by the second degree polynomial in generators $J_n^{+,0,-}$, which can be explicitly written in the form,
\begin{equation}
\label{P2Q}
 \mathcal{P}_2({\sf L}_0,\ag)\ =\  c_{2,0} {\sf L}^2_0 + {\sf L}_0 (c_{1,1} \ag + c_{1,0}) + c_{0,2} \ag^2 + c_{0,1} \ag + c_{0,0} \ ,
\end{equation}
or, by substituting ${\sf L}_0\ =\ \bg \ag $, equivalently, 
\begin{equation}
\label{P2Qba}
\mathcal{Q}_2 (\bg,\ag)\ =\ Q_4(\bg)\, \ag^2\ +\ Q_3(\bg)\, \ag\ +\ Q_2(\bg)\ ,
\end{equation}
where $Q_{4,3,2}(\bg)$ are polynomials of degrees 4,3,2 in $\bg$, respectively,
\[
    Q_4 = a_{4}\,\bg^4 + a_{3}\,\bg^3 + a_{2}\,\bg^2 + a_{1}\,\bg + a_{0} \ ,
\]
\[
    Q_3 = b_{3}\,\bg^3 + b_{2}\,\bg^2 + b_{1}\,\bg + b_{0} \ ,
\]
\[
    Q_2 = d_{2}\,\bg^2 + d_{1}\,\bg + d_{0} \ ,
\]
where $\{a_j,b_j,d_j\}$ are complex parameters with two constraints,
\begin{equation}
\label{constraint-1}
  a_{4} n (n-1) + b_{3} n + d_2\ =\ 0 \ ,
\end{equation}
\begin{equation}
\label{constraint-2}
  a_{4} (n-1) (n-2) + b_{3} (n-1) + d_2 + a_{3} n (n-1) + b_{2} n + d_1\ =\ 0 \ .
\end{equation}
These constraints guarantee that $\mathcal{P}_2 (\bg,\ag)$ has a finite-dimensional invariant subspace $\mathcal{L}_n$. Without loss of generality the parameters $a_0, d_0$ can be set equal to zero, $a_0=d_0=0$. 

In the coordinate-momentum representation on the real line $\ag = \frac{d}{dx}$ and $\bg= x$ (\ref{c-m-reps})
the operator $\mathcal{P}_2 (\bg,\ag)$ becomes the differential operator,
\begin{equation}
\label{P2Qba-x}
 \mathcal{P}_2 (x,\frac{d}{dx})\ =\ Q_4(x)\,\frac{d^2}{dx^2}\ +\ Q_3(x)\, \frac{d}{dx}\ +\ Q_2(x)\ ,
\end{equation}
with Fuchs index $n_f=2$ 
\footnote{For a linear differential operator $L(x, \pa_x )$ with polynomial coefficients the Fuchs index $n_f$ is defined as follows: $L: x^n \rar P_{n+n_f}(x)$. For the hypergeometric operator $n_f=0$. }. 
The operator (\ref{P2Qba-x}) has a finite-dimensional invariant subspace consisting of the polynomials
of degree $n$: $P_n=\langle x^i\,|\, 0 \leq i \leq n \rangle$. It has $(n+1)$ polynomial eigenfunctions with eigenvalues given by roots of an algebraic equation of $(n+1)$th degree.

In the complex $(z,\overline{z})$ representation (\ref{repr-in-z}) the operators (\ref{P2Q}), (\ref{P2Qba}) become sixth order differential operators. All $(n+1)$ eigenspaces are 
of infinite multiplicity, they are elements of the space $\mathcal{L}_n$, see (\ref{curly L}), 
and are given by $n$-polyanalytic functions. Their eigenvalues 
coincide with those of the operator $\mathcal{P}_2 (x,\frac{d}{dx})$ (\ref{P2Qba-x}).

\subsection{Heun operator in the Fock space}

By setting $c_{2,0}=0$ in (\ref{P2Q}) or, equivalently, $a_4=b_3=d_2=0$ in (\ref{P2Qba}) we arrive at 
\begin{equation}
\label{P2Q-3}
 \mathcal{P}_2({\sf L}_0,\ag)\ =\ {\sf L}_0 (c_{1,1} \ag + c_{1,0}) + c_{0,2} \ag^2 + c_{0,1} \ag + c_{0,0} \ ,
\end{equation}
or, by substituting ${\sf L}_0\ =\ \bg \ag $, equivalently,
\begin{equation}
\label{P2Qba-3}
  \mathcal{Q}_2 (\bg,\ag)\ =\ {\hat Q}_3(\bg)\, \ag^2\ +\ {\hat Q}_2(\bg)\, \ag\ +\ {\hat Q}_1(\bg)\ ,
\end{equation}
where ${\hat Q}_{3,2,1}(\bg)$ are polynomials of degrees 3,2,1 in $\bg$, respectively, 
\[
    \hat Q_3 = a_{3}\,\bg^3 + a_{2}\,\bg^2 + a_{1}\,\bg  \ ,
\]
\[
    \hat Q_2 =  b_{2}\,\bg^2 + b_{1}\,\bg + b_{0} \ ,
\]
\[
    \hat Q_1 = d_{1}\,\bg  \ ,
\]
with the single imposed constraint 
\begin{equation}
\label{constraint-heun}
a_{3} n (n-1) + b_{2} n + d_1\ =\ 0 \ ,
\end{equation}
cf. (\ref{constraint-2}). This constraint guarantees that $\mathcal{P}_2 (\bg,\ag)$ has a finite-dimensional invariant subspace $\mathcal{L}_n$, see (3.13). We call the operator $\mathcal{P}_2 (\bg,\ag)$ {\it the Heun operator in the Fock space}.

In the coordinate-momentum representation on the real line $\ag = \frac{d}{dx}$ and $\bg= x$ (\ref{c-m-reps}), the operator $\mathcal{\hat P}_2 (\bg,\ag)$ becomes the differential operator,
\begin{equation}
\label{P2Qba-x-3}
 \mathcal{P}_2 (x,\frac{d}{dx})\ =\ {\hat Q}_3(x)\,\frac{d^2}{dx^2}\ +\ {\hat Q}_2(x)\, \frac{d}{dx}\ +\ {\hat Q}_1(x)\ ,
\end{equation}
with Fuchs index $n_f=1$. With the imposed constraint (\ref{constraint-heun}), this operator coincides with celebrated Heun operator \cite{Turbiner:2016}. 
The operator (\ref{P2Qba-x-3}) has a finite-dimensional invariant subspace consisting of the polynomials of degree $n$: $P_n=\langle x^i\,|\, 0 \leq i \leq n \rangle$. It has 
$(n+1)$ polynomial eigenfunctions with eigenvalues given by roots of a characteristic algebraic equation of $(n+1)$th degree. All known (quasi)-exactly solvable problems presented in \cite{Turbiner:1988} in the algebraic form \footnote{For review see \cite{Turbiner:2016-PR}} 
are particular cases of operator (\ref{P2Qba-x-3}), see for details \cite{Turbiner:2016}. 
An outstanding property of these operators is that the first $(n+1)$ eigenfunctions are orthogonal polynomials of degree $n$, which can be found by algebraic means, while the remaining eigenfunctions 
are of non-polynomial nature. 

Note that by taking the operators $\ag_{\delta}$ and $\bg_{\delta}$ of Example \ref{ex:delta}, 
see (2.5), or the operators $\ag_q$ and $\bg_q$ of Example \ref{ex:q} , see (2.6), 
instead of $\ag$ and $\bg$ in the operator (\ref{P2Qba-3}), we arrive at the finite-difference 
or discrete operators acting on the uniform or exponential lattices with $(n+1)$ polynomial eigenfunctions. Remarkably, the corresponding eigenvalues of 
(\ref{P2Qba-3}) and (\ref{P2Qba-x-3}) and of those finite-difference or discrete operators {\it coincide}. Hence, all four operators are isospectral in the subset of polynomial eigenfunctions, respectively.

In the complex $(z,\overline{z})$ representation (\ref{repr-in-z}), the operators (\ref{P2Q-3}), (\ref{P2Qba-3}) become fifth order differential operators. The $(n+1)$ eigenspaces are of infinite multiplicity, they are elements of the space $\mathcal{L}_n$ (3.13) and are given by $n$-polyanalytic functions. Their eigenvalues coincide with those of the operator $\mathcal{P}_2 (x,\frac{d}{dx})$ (\ref{P2Qba-x-3}).

\subsection{Lam\'e operator in the Fock space}

Let us take the operator  
\begin{equation}
\label{lame}
   {\bf L} (\bg,\ag)\ = \ 4({\bg}^3 - 3 \, \la {\bg}^2\ +\ 3 \de {\bg}) \ag^2 \ +\
    6 ({\bg}^2-2 \la {\bg}+\de) \ag\ -\ 2n(2n+1) ({\bg} - \la)\ ,
\end{equation}
where $\la, \de, n$ are some real parameters. This operator is a particular case of the Heun operator $\mathcal{\hat P}_2$ (\ref{P2Qba-3}). If $n$ is a non-negative integer, the operator 
has a finite-dimensional invariant subspace $\mathcal{L}_n$, see (\ref{curly L}). 
We call the operator ${\bf L} (\bg,\ag)$ {\it the Lam\'e operator in the Fock space}.

In the coordinate-momentum representation on the real line $\ag = \pa_x$ and $\bg= x$ (\ref{c-m-reps})
the operator ${\bf L} (\bg,\ag)$ becomes the differential operator,
\begin{equation}
\label{lame-x}
    {\bf L} (x, \pa_x)\ = \ 4({x}^3 - 3 \, \la {x}^2\ +\ 3 \de {x}) \pa^2_{{x}} \ +\
    6 ({x}^2-2 \la {x}+\de) \pa_{{x}}\ -\ 2n(2n+1) ({x} - \la)\ ,
\end{equation}
which can be immediately recognized as the celebrated Lam\'e operator in algebraic form, see e.g. \cite{Turbiner:2016-PR}. The Lam\'e operator has $(n+1)$ polynomial eigenfunctions in the form 
of polynomials in $x$ of degree $n$.

By introducing the new variable $x=\wp(\ta)$, where $\wp (\ta)\equiv \wp(\ta|g_2,g_3)$ is the $\wp$-Weierstrass function and the parameters $\la, \de$ are related 
to the elliptic invariants,
\[
     g_2 = 12 (\la^2 - \de)\ ,\qquad  g_3=4 \la (2\la^2-3\de)\ ,
\]
we arrive at the self-adjoint form of the operator ${\bf L}$,
\begin{equation}
\label{Lame-H}
    {\mathcal H}(\ta)\ \equiv \ {\bf L} (\ta, \pa_{\ta})\ =\ - \pa^2_{{\ta}} \ +\ (2n+1) 2n\ \wp (\ta) \ ,
\end{equation}
which is called the Lam\'e Hamiltonian. Since the potential $\wp$ is chosen to be periodic 
with one real and one pure imaginary periods, ${\mathcal H}$ is characterized 
by a zone structure and the eigenvalues, which correspond to polynomial eigenfunctions, 
describe the lower edges of permitted zones.

By substituting the operators $\ag_{\delta}$ and $\bg_{\delta}$ see (2.5), 
or the operators $\ag_q$ and $\bg_q$ , see (2.6), instead of $\ag$ and $\bg$ into the operator (\ref{lame}), we arrive at the finite-difference or discrete operators, acting on the uniform or exponential lattices, with $(n+1)$ polynomial in $x$ eigenfunctions. Remarkably, the corresponding eigenvalues of those finite-difference or discrete operators {\it coincide} with those of the operator (\ref{lame}) or (\ref{lame-x}). Hence, all four operators are isospectral in the sector of polynomial eigenfunctions, respectively. 

In the complex $(z,\overline{z})$ representation (\ref{repr-in-z}), the operator (\ref{lame}) becomes a fifth order differential operator. The $(n+1)$ eigenspaces are 
of infinite multiplicity, they are subspaces of the space $\mathcal{L}_n$ and 
are given by $n$-poly-analytic functions. Their eigenvalues coincide with those 
of the operator (\ref{lame-x}) and (\ref{Lame-H}).

\subsection{QES problem with sextic polynomial potential}

Let us take the operator  
\begin{equation}
\label{sextic}
   {\bf h}_6 (\bg,\ag)\ = \ -4 {\bg} \ag^2 \ +\
    2(2a {\bg}^2+2 b {\bg} -1) \ag\ -\ 4 a n {\bg} \ ,
\end{equation}
where $a, b, n$ are the real parameters. This operator is a particular case of the Heun operator
$\mathcal{\hat P}_2$ (\ref{P2Qba-3}). If $n$ is a non-negative integer, it has a finite-dimensional 
invariant subspace $\mathcal{L}_n$, see (\ref{curly L}). We call the operator ${\bf h}_6 (\bg,\ag)$ 
{\it the sextic QES polynomial potential operator in the Fock space}, or {\it the sextic QES operator in the Fock space}, for brevity. 

In the coordinate-momentum representation on the real line, $\ag = \pa_x$ and $\bg= x$ (\ref{c-m-reps}), the operator ${\bf h}_6 (\bg,\ag)$ (\ref{sextic}) becomes the differential operator,
\begin{equation}
\label{sextic-x}
    {\bf h}_6 (x, \pa_x)\ = \ - 4 {x} \pa_{x}^2\ +\ 
    2(2a x^2+2 b x -1) \pa_{x}\ -\ 4 a n x \ ,
\end{equation}
see \cite{Turbiner:1988,Turbiner:2016-PR}, where $a,b$ are free parameters,
which can be immediately recognized as the celebrated sextic polynomial 
QES oscillator in algebraic form, 
see e.g. \cite{Turbiner:2016-PR}. This operator has $(n+1)$ polynomial eigenfunctions 
in the form of polynomials of degree $n$, which are orthogonal with the weight factor 
$\psi_0(x)=e^{-a x^2 - b x}$.
By making the gauge rotation $(\psi_0)^{-1}\, {\bf h}_6 (x, \pa_x)\, \psi_0$ and changing variable, $x=\ta^2$ 
we arrive at the Hamiltonian of the sextic polynomial QES oscillator,
\begin{equation}
\label{sextic-tau}
    {\bf H}_6 (\ta, \pa_{\ta})\ = \ - \pa_{\ta}^2\ +\ 
    a^2 \ta^6 + 2 ab \ta^4 + [b^2 - (4n + 3) a] x^2 - b \ ,
\end{equation}
widely used in applications. 

Remarkably, in the complex $(z,\overline{z})$ representation (\ref{repr-in-z}), 
the operator (\ref{sextic}) becomes a third order differential operator. 
The $(n+1)$ eigenspaces are of infinite multiplicity, they are elements of 
the space $\mathcal{L}_n$ (3.13) and are given by $n$-poly-analytic functions. 
Their eigenvalues can be found algebraically by solving a characteristic (polynomial) 
equation of degree $(n+1)$, they coincide with those of the operator (\ref{sextic}) and (\ref{sextic-x})-(\ref{sextic-tau}).

\vskip 1cm

\section{Conclusions}

The $k$-polyanalytic functions were introduced as elements of the kernel of the 
$k$th degree of the Cauchy-Riemann 
operator 
\begin{equation*}
 \left(\frac{\pa}{\pa \overline{z}}\right)^k f = 0\ ,\ k \in \mathbb{N}\ ,
\end{equation*}
see e.g. \cite{Balk:1991,Abreu:2014}. In our paper \cite{TV} we showed that the space of $k$-polyanalytic functions coincides with the finite-dimensional representation space of the $\mathfrak{sl}(2)$-algebra (\ref{sl2ag}) in the complex $(z,\overline{z})$ representation 
(\ref{repr-in-z}). In this paper we present a $\mathfrak{sl}(2)$-algebraic classification 
of the linear differential operators in $(\frac{\pa}{\pa \overline{z}}\,,\,\frac{\pa}{\pa z})$ 
having the space of poly-analytic functions as an invariant subspace. A fundamental role in this classification is played the number operator or, equivalently, the Euler-Cartan operator,

\[
   L_0\ =\ \bg \ag\ =\ \frac{\pa^2}{\pa z\,\pa\overline{z}} + 
   \overline{z}\,\frac{\pa}{\pa\overline{z}}\ ,
\]
see (\ref{landau}): it maps the space of $k$-polyanalytic functions to itself 
\footnote{Note the Euler-Cartan operator is defined up to an additive constant}.
By taking this operator as the Cartan generator, the algebra $\mathfrak{sl}(2)$ 
in the representation (\ref{sl2ag-L0}) can be uniquely reconstructed under 
the requirement that the Gauss decomposition holds: 
the Lie bracket of the raising and lowering generators must be equal 
to the Cartan generator. A similar procedure allows a systematic definition of  
the polyanalytic functions of two and more complex variables. It is worth noting that
for the case of two variables the number operator has the form,
\[
   L_0\ =\ \bg_1 \ag_1 + k \bg_2 \ag_2 \  ,
\]
where $k$ is a non-negative integer, $k=1,2,3, \ldots $. This operator is a Cartan generator
of infinite-dimensional, $(2k+6)$-generated algebra $g^{(k)}$, described in \cite{ST:2013}.

\vskip 1cm

\section*{My friend Kolia Vasilevski: personal recollections by AVT}

Mikhail (Misha) A Shubin - a renowned Soviet-American mathematician and 
lifelong personal friend of AVT - introduced Nikolai (Kolia) L Vasilevski and 
myself in 1988 in Moscow (FSU), when all three of us were studying together 
spoken English. In 1994, when I moved to Mexico together with my family, 
Misha Shubin informed me that Kolia and his family were already living in Mexico-city 
and might help us establish ourselves in this country. It was a vital piece of information 
but to our great surprise we learned that the distance between CINVESTAV, where Kolia worked, 
and UNAM, which offered me a position, is about 30km. Thus, two different universities, 
two different groups of people, but the same culture. As a result, we travelled this distances 
``infinitely"-many times visiting each other and talked over the phone almost every day for almost 30 years! In fact, we shared our life by helping each other understand what is happening 
around us, also sharing our love to mountains, and to the Mexican nature and 
cuisine. It is hard to give a list of the places we visited together all over the country. 
Our children were more or less of the same age, they shared life as well. Being Odessite, 
of the Odessa, Ukraine origin, Kolia had a fine sense of humor (which was sometimes quite disturbing) and loved to joke, it simplified our life enormously: the first ten years 
in Mexico were extremely difficult for all of us.

In 1998 we went with Kolia to Ajusco near Mexico-City to summit the Paseo del Marques - a peak 
about 4000m above sea level - to celebrate 
Kolia's 50th birthday: Kolia brought a small bottle of the fine French red wine, and, in turn, 
I brought fancy glasses and we were drinking at 4000 meters altitude to Kolia's health and prosperity. Perhaps, for the first time Kolia started to talk about mathematics 
(we had a tacit agreement not to talk about science in front of our wives and children): 
he told me about poly-analytic functions. I was fascinated by this simple and natural idea, 
and responded promptly saying there must exist a Lie-algebraic interpretation: 
poly-analytic functions must be related to representation spaces of finite-dimensional representations 
of the algebra $sl_2$. Kolia asked me to write a ``two pages of summary" 
to understand what was I talking about.
It took two years to produce a type of summary: in 2000 the first version 
of our future joint paper was written. It contained about 4 pages.
I could not imagine that it would take about 20 years to prepare the final, 42nd version of the text: 
every sentence was discussed carefully! 
These twenty years of joint work went through the tough process of mutual education: 
Kolia taught me constructive advanced complex (and functional) analysis, while I taught him constructive 
representation theory of the Lie algebras (of differential operators) and as well as ODE/PDE. We prepared 
homeworks for each other which were not always well received and positively evaluated by another party: 
the differences between the mathematical and physical mentalities/cultures were conceptual but in spite 
of that we did our best to overtake them, 
to understand each other. I was deeply impressed how meticulous Kolia was! Finding 
even a small error/mistake in his own results led him to a kind of tragedy, I tried to cool 
him down. It was clear that he was not used to making mistakes, in these moments he lost 
his sense of humor completely:
I never knew {\it such} Kolia in regular, non-scientific life. But a moment arrived when 
we became unable to reach an agreement. 
We made a decision to prepare two separate papers, mathematical and physical. 
In the former paper, the final word was by Kolia, while in the latter one the final 
word was mine, no discussions. Only under such an agreement we were able to complete and successfully 
publish the ``mathematical" paper \cite{TV} 
\footnote{Developing the mathematical side of this paper in 2021-2023 
Kolia published three papers on the subject: 
N.L.~Vasilevski, {\it COMPLEX ANALYSIS AND OPERATOR THEORY, \bf 16(5)}, 2022; 
ibid {\bf 17(6)}, 2023; {\it BOLETIN DE LA SOCIEDAD MATEMATICA MEXICANA, \bf 29}, 2023, 97}. 
The ``physical" paper is presented 
here - originally it was supposed to have no proofs, thus, not to be rigorous mathematically, 
and to be published in a physics-oriented journal. 
It is unfinished in the sense that the 10th draft of the most important Sections 2-3 
was prepared by Kolia (thus, with his final word) - it took about 3 years of intense 
discussions - and we agreed to meet to make this draft into its final form. 
Two days before our meeting Kolia unexpectedly passed away. But five days before 
the meeting Kolia phoned me - I was visiting Stony Brook University at that 
moment - to remind (and confirm) the meeting. 
Preparing the manuscript I made the decision to keep the Kolia's final draft of 
Sections 2-3 maximally unchanged, just making small edits, as a memory to my dear 
lifelong friend and co-author. Introduction and Conclusions are added as well as Section 4 
following our preliminary notes. I have no plans to continue this scientific direction 
in Kolia's absence. More than one year has passed from Kolia's departure, 
but still I can't afford it.\ R.I.P.

\end{document}